\documentclass[preprint,showpacs,showkeys]{revtex4}
\usepackage{bm}
\usepackage{graphicx}
\usepackage{amsmath, amsthm, amsfonts,amssymb}

\begin{document}

\title{Measuring Volatility Clustering in Stock Markets}

\author{Gabjin Oh}
\email{gq478051@postech.ac.kr}
\affiliation{NCSL, Department of Physics, Pohang University of Science and Technology, Pohang, Gyeongbuk, 790-784, Korea}
\affiliation{Asia Pacific Center for Theoretical Physics, Pohang, Gyeongbuk, 790-784, Korea}

\author{Cheoljun Eom}
\email{shunter@pusan.ac.kr}
\affiliation{Division of Business Administration, Pusan National University, Busan 609-735, Korea}

\author{Seunghwan Kim}
\email{swan@postech.ac.kr}
\affiliation{NCSL, Department of Physics, Pohang University of Science and Technology, Pohang, Gyeongbuk, 790-784, Korea}
\affiliation{Asia Pacific Center for Theoretical Physics, Pohang, Gyeongbuk, 790-784, Korea}

\author{Taehyuk Kim}
\email{tahykim@pusan.ac.kr}
\affiliation{Division of Business Administration, Pusan National University, Busan 609-735, Korea}

\begin{abstract}
We propose a novel method to quantify the clustering behavior in a
complex time series and apply it to a high-frequency data of the
financial markets. We find that regardless of used data sets, all
data exhibits the volatility clustering properties, whereas those
which filtered the volatility clustering effect by using the GARCH
model reduce volatility clustering significantly. The result
confirms that our method can measure the volatility clustering
effect in financial market.
\end{abstract}

\maketitle

\section{Introduction}
Recently, financial markets have been known as representatives of
complex system, which changes the property of system dynamically
according to inflows of various information from outside and
interactions between heterogenous agents [1]. In order to
understand the complexity of financial market, the methods of
interdisciplinary research have been achieving in physics and
economics fields. The various stylized facts such as  the
long-term memory of volatility [2], volatility clustering [3], fat
tails [4], multifractality [5, 6] are observed. The obvious
properties among the stylized facts are the long-term memory
property and clustering effects of the volatility data. In
previous studies, clustering behaviors are shown in the return
time interval statistics of the climate records [7], medical data,
extreme floods [8], and economics [9].

The various models which reflects the volatility clustering effect
in order to predict exactly the volatility in the econometrics
field are introduced. The autoregressive conditional
heteroskedasticity (ARCH) [10] and the generalized autoregressive
conditional heteroskedasticity (GARCH) model [11] are the
representatives. Namely, the many researches to understand the
micro-mechanism of market has been processed. However, the study
to quantify the volatility clustering effects is not sufficient
yet. If we observe quantitatively the volatility clustering effect
in financial markets, we will understand micro phenomena of the
market. In this paper, we propose the novel method to quantify the
volatility clustering effect in the financial time series.

We find that all data sets analyzed in this paper exhibit the volatility clustering property, whereas the data which filters the volatility clustering effect by the GARCH(1,1) model reduces the degree of volatility clustering significantly.

In the next section, we describe the data sets and methods used in this paper. In section \ref{sec:RESULTS}, we preset our results of this study. Section \ref{sec:CONCLUSIONS} concludes.

\section{DATA and METHODS}

\subsection{DATA}

We investigate quantitatively the volatility clustering behaviors using financial time series including the following market data sets: the 5 minute S\&P 500 index from 1995 to 2004 and the 5 minute 28 individual stocks traded in the NYSE with the largest liquidity from 1993 to 2002. The return time series $r(t)$ is calculated by the
log-difference of high-frequency prices as follows: $r(t) = \ln P(t) -
\ln P(t-1)$ where $P(t)$ represents the stock price at time $t$.

\subsection{Method to quantify the volatility clustering}

In this subsection, we propose a novel method to quantify the
volatility clustering effect. We estimate and analyze quantitatively the volatility clustering
effect existed in the financial time series. The process is explained by the following.

\textbf{Step 1} (The symbolized process): We transfer the return
time series $r(t)$ to the symbolic data $s(t)$ in order to
quantify the volatility clustering effect in the financial data
using the control parameter, such as the number of bins which is
defined as

\begin{equation}
\{ S(t)=T_{i},  ~~~~~if  ~~ r(t) \in T_{i} \}, ~~ T_{i} = \{T_{1}, T_{2}, \cdots, T_{N_{b}}\}
\end{equation}
where $N_b$ is the number of bins. The conditional distribution with statistical significance is calculated by the symbolized process.

\textbf{Step 2} (Calculating the conditional distribution): We estimate the conditional distribution using the symbolized time series generated in step 1. In other words, the conditional distribution corresponds to the next value of a specific symbol $S_T$ in the symbolic data. Next, we calculate repeatedly the conditional distribution $P(S_j | S_T)$ for all symbolic data in the proper regime. The conditional distribution of each symbolic data has a non-trivial property like the conditional value $S_T$ if there is a volatility clustering behavior.

\textbf{Step 3} (The average value of conditional distribution): The step 3 is the calculation of the average value of the conditional distribution estimated in the step 2. We only consider the conditional distribution of symbolic data in the proper range because the extreme symbolic data are rare. By the average value of the conditional distribution, we observe the volatility clustering effect defined as

\begin{equation}
\overline {S_{T}} = \frac{1}{N_T} \sum_{j=1}^{N_T}|P(S_j | S_T)|
\end{equation}
where $N_T$ is the element numbers of the conditional distribution
$P(S_j | S_T)$ in terms of a specific symbol $S_T$. If the average
value is not dependent on the symbolic value $S_T$, there is no
volatility clustering effect because the time series shows the
volatility clustering effect only when it has a positive
(negative) relation with the positive (negative) values of
$S_{T}$. Next, we calculate the relation between the specific
symbolic values $S_{T}$ and average values $\overline{S_{T}}$ of
conditional distribution. In other words, we observe the degree of
volatility clustering (DVC) behavior according to the relationship
between $\overline{S_{T}}$ and $S_{T}$. The average value
$\overline{S_{T}}$ is definded as

\begin{displaymath}
\overline{S_T} = \left\{
\begin{array}{ll}
DVC^{P} \times S_T &S_t \geq 0\\
DVC^{N} \times S_T &\textrm{otherwise}\\
\end{array}
\right.
\end{displaymath}
where $DVC^{P,N}$ is the degree of volatility clustering effect
for the positive and negative cases respectively. When $DVC^{P,N}
= 0$, there is no clustering effect. However, when the value of
$DVC^{P,N}$ is nonzero, the degree of volatility clustering effect
according to the relative magnitude of $DVC^{P,N}$ is measured.
Therefore, we can estimate quantitatively the volatility
clustering effect of the financial time series.

\section{Results}
\label{sec:RESULTS}

In this section, we present the volatility clustering effect of the financial time series. In order to verify usefulness of the method proposed in this paper, we employ the GARCH(1,1) which reflects the volatility clustering effect.

First of all, we apply the novel method to the 5 minute S\&P500 index and calculate the degree of volatility clustering. Fig. 1a represents the return time series of the 5 minute S\&P500 index and Fig. 1b shows its symbolic time series. We then calculate the conditional distribution $P(S|S_T)$ of a specific symbol data $S_T$. Fig. 1c shows the conditional distributions of specific symbols. In Fig. 1c, we find that the width of the conditional distribution increases as the value of symbolic data increases. In other words, the width of the conditional distribution of small symbolic data is relatively narrow than that of large symbolic data. The average value for the conditional distribution regarding specific symbolic data $S_T$ is calculated in order to observe the relationship between  specific symbolic values and its conditional distribution. Fig. 1d shows the relationship between specific symbol and average value. Circles and squares of Fig. 1d indicate the original and the surrogate time series respectively. We find that the average values of the conditional distribution for the original time series are positively related to the magnitude of specific symbolic value, $DVC_{S\&P500}^{P}=0.57$ and $DVC_{S\&P500}^{N}=0.52$, while those for the shuffled time series is not dependent on the symbolic value $S_{T}$. The return time series of the S\&P500 index shows the volatility clustering effect, the larger (small) values follow the larger (small) values.

Next, we utilize the GARCH model to verify the usefulness of our
method. The GARCH model generates the volatility clustering
effect. We create the new time series with the volatility
clustering effect removed by the GARCH(1,1) filtering model and
estimate the degree of the volatility clustering effect. Fig. 2
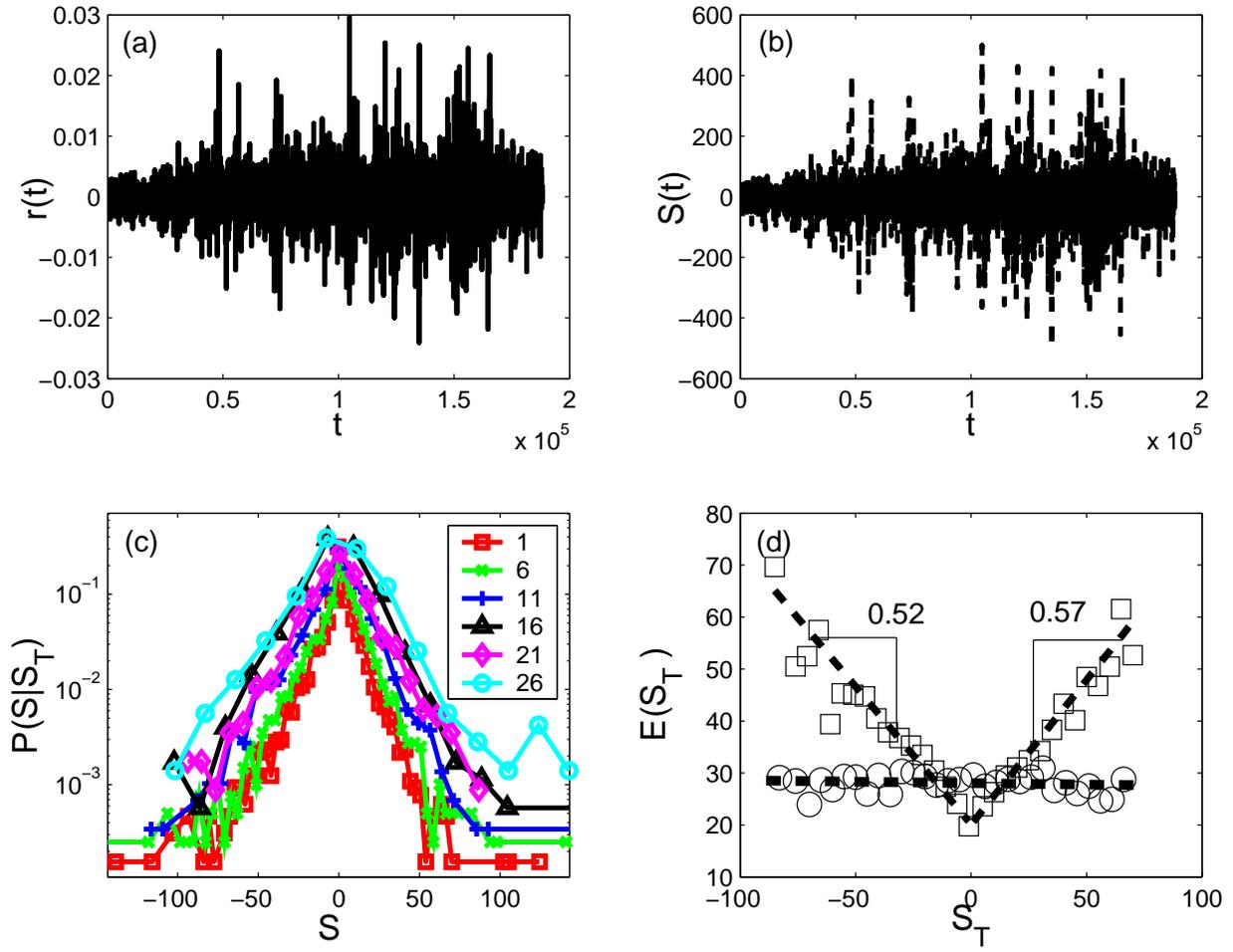
displays the degree of the volatility clustering for the 28
individual stocks traded in the NYSE stock market with the largest
liquidity. The circles (red), the diamonds (blue), the squares
(green), and the triangles (pink) indicate the degree of the
volatility clustering for the positive and negative return time
series using the original and the GARCH(1,1) filtering data,
respectively. In Fig. 2, we find that all the stock return time
series, regardless of individual stocks, have the volatility
clustering effect, $0.38 \leq DVC^P \leq 0.72$ and $-0.69 \leq
DVC^N \leq -0.32$. However, after eliminating the volatility
clustering behavior by the GARCH(1,1) model, the degree of the
volatility clustering effect is reduced significantly. This
supports that our method to quantify the volatility clustering
effect in financial time series is working well.

\section{Conclusions}
\label{sec:CONCLUSIONS}

We proposed the novel method to quantify the volatility clustering
behavior in financial time series and calculated the degree of the
volatility clustering (DVC) using the diverse stock prices. First,
we found that all financial data analyzed exhibited the volatility
clustering properties, whereas those which are filtered the
volatility clustering effect by the GARCH(1,1) model reduced the
degree of the volatility clustering effect significantly. This
result confirmed that our method calculated the volatility
clustering effect in financial time series well. Our method might
be applied to elaborate clustering analysis of diverse complex
signals including climate, HRV as well as financial time series.
Further studies on the volatility clustering will examine to the
above systems more extensively.

This work was supported by the Korea Research Foundation funded by
the Korean Government (MOEHRD) (KRF-2005-042-B00075), and the
MOST/KOSEF to the National Core Research Center for Systems
Bio-Dynamics (R15-2004-033), and by the Ministry of Science \&
Technology through the National Research Laboratory Project, and
by the Ministry of Education through the program BK 21.

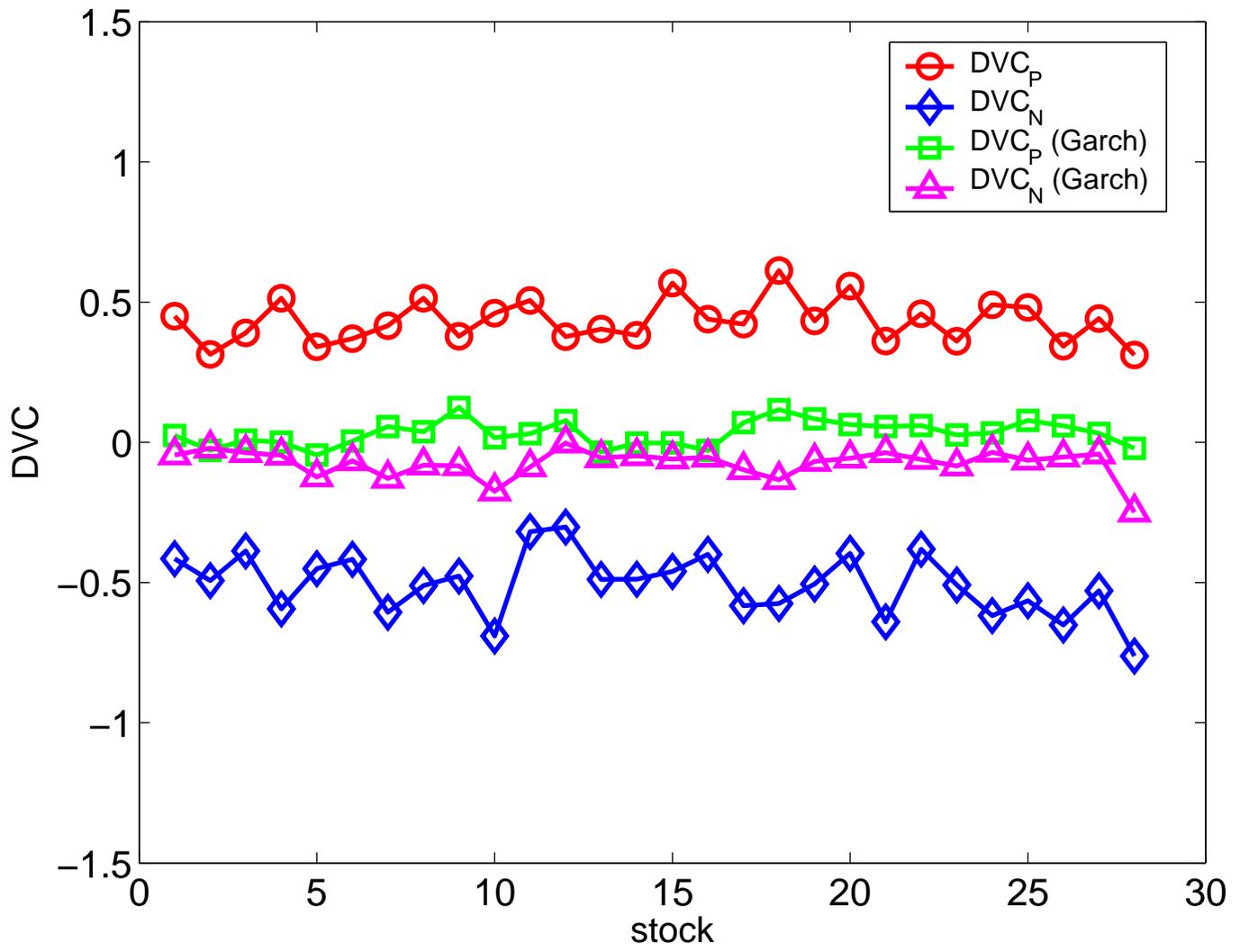
\end{document}